\documentclass[aps,prb,twocolumn,showpacs]{revtex4}

\usepackage{graphicx}
\usepackage{amsmath}
\usepackage{amssymb}

\begin{document}

\title{Interplay between charge order and superconductivity in cuprate superconductors}

\author{Deheng Gao and Yiqun Liu}

\affiliation{Department of Physics, Beijing Normal University, Beijing 100875, China}

\author{Huaisong Zhao}

\affiliation{College of Physics, Qingdao University, Qingdao 266071, China}

\author{Yingping Mou and Shiping Feng}
\email{spfeng@bnu.edu.cn}

\affiliation{Department of Physics, Beijing Normal University, Beijing 100875, China ~}

\begin{abstract}
One of the central issues in the recent study of cuprate superconductors is the interplay of charge order with superconductivity. Here the interplay of charge order with superconductivity in cuprate superconductors is studied based on the kinetic-energy-driven superconducting (SC) mechanism by taking into account the intertwining between the pseudogap and SC gap. It is shown that the appearance of the Fermi pockets is closely associated with the emergence of the pseudogap. However, the distribution of the spectral weight of the SC-state quasiparticle spectrum on the Fermi arc, or equivalently the front side of the Fermi pocket, and back side of Fermi pocket is extremely anisotropic, where the most part of the spectral weight is located around the tips of the Fermi arcs, which in this case coincide with the hot spots on the electron Fermi surface (EFS). In particular, as charge order in the normal-state, this EFS instability drives charge order in the SC-state, with the charge-order wave vector that is well consistent with the wave vector connecting the hot spots on the straight Fermi arcs. Furthermore, this charge-order state is doping dependent, with the charge-order wave vector that decreases in magnitude with the increase of doping. Although there is a coexistence of charge order and superconductivity, this charge order antagonizes superconductivity. The results from the SC-state dynamical charge structure factor indicate the existence of a quantitative connection between the low-energy electronic structure and collective response of the electron density. The theory also shows that the pseudogap and charge order have a root in common, they and superconductivity are a natural consequence of the strong electron correlation.
\end{abstract}


\maketitle

\section{Introduction}

The understanding of the mechanism of superconductivity in cuprate superconductors remains one of the most intriguing problems in condensed matter physics. The parent compound of cuprate superconductors is a half-filled Mott insulator \cite{Bednorz86,Kastner98}, which occurs to be due to the strong electron correlation \cite{Anderson87,Phillips10}. Superconductivity is derived from doping this parent Mott insulator \cite{Bednorz86,Kastner98}, indicating that superconductivity and the related exotic physics in the doped regime are also dominated by the same strong electron correlation. In conventional superconductors \cite{Bardeen57,Schrieffer64}, an energy gap exists in the electronic energy spectrum only below the superconducing (SC) transition temperature $T_{\rm c}$, which is corresponding to the energy for breaking a Cooper pair of the electrons and creating two excited states. However, in cuprate superconductors above $T_{\rm c}$ but below a characteristic temperature $T^{*}$, an energy gap called the pseudogap exists \cite{Timusk99,Hufner08}. In particular, this pseudogap is most notorious in the underdoped regime, where the charge carrier concentration is too low for the optimal superconductivity \cite{Timusk99,Hufner08}.

However, the strong electron correlation also induces the system to find new way to lower its total energy, often by spontaneous breaking of the native symmetries of the lattice \cite{Comin16}. This tendency leads to that the pseudogap regime harbours diverse manifestations of the ordered electronic phases, and then a characteristic feature in the complicated phase diagram of cuprate superconductors is the interplay between different ordered electronic states and superconductivity \cite{Timusk99,Hufner08,Comin16}. In particular, by virtue of systematic studies using the scanning tunneling microscopy (STM), resonant X-ray scattering (RXS), angle-resolved photoemission spectroscopy (ARPES), and many other measurement technique \cite{Comin16,Comin14,Wu11,Chang12,Ghiringhelli12,Neto14,Fujita14,Croft14,Hucker14,Campi15,Comin15a,Hashimoto15,Peng16,Hinton16}, it has been found recently that charge order is a universal phenomenon in cuprate superconductors, which exists within the pseudogap phase, appearing below a temperature $T_{\rm CO}$ well above $T_{\rm c}$ in the underdoped regime, and coexists with superconductivity below $T_{\rm c}$. $T_{\rm CO}$ is the temperature where charge order develops, and is of the order of the pseudogap crossover temperature $T^{*}$. This near coincidence of $T^{*}$ and $T_{\rm CO}$, as well as the coexistence of charge order and superconductivity below $T_{\rm c}$, suggests that a crucial role in the pseodogap phase is played by charge order \cite{Comin16}. These experimental observations also identified that charge order in the pseudogap phase of cuprate superconductors emergences consistently in surface and bulk, and in momentum and real space. Furthermore, the combination of the RXS data and electron Fermi surface (EFS) measured results using ARPES revealed a quantitative link between the charge-order wave vector $Q_{\rm CO}$ and the momentum vector connecting the tips of the straight Fermi arcs \cite{Comin16,Comin14,Neto14}, which in this case coincide with the hot spots on EFS, indicating that the hot spots play an important role in the charge-order formation. This correspondence also shows the existence of a quantitative connection between the collective response of the electron density and the low-energy electronic structure. As a natural consequence of a doped Mott insulator, the charge-order state is also doping dependent, with the magnitude of the charge-order wave vector $Q_{\rm CO}$ that decreases upon the increase of doping, in the analogy of the unusual behavior of the doping dependence of the pseudogap \cite{Comin16,Comin14,Wu11,Chang12,Ghiringhelli12,Neto14,Fujita14,Croft14,Hucker14,Campi15,Comin15a,Hashimoto15,Peng16,Hinton16}. These experimental results observed on cuprate superconductors \cite{Comin16,Comin14,Wu11,Chang12,Ghiringhelli12,Neto14,Fujita14,Croft14,Hucker14,Campi15,Comin15a,Hashimoto15,Peng16,Hinton16} therefore show that charge order more intrinsically intertwines with superconductivity. In this case, some crucial questions are raised: (i) does the strong electron correlation play a role in the charge-order state and its interplay with superconductivity? (ii) is charge order also the result of the emergence of the pseudogap? (iii) do charge order and the SC order compete?

Since the discovery of charge order and its evolution with doping and temperature in the pseudogap phase of cuprate superconductors, the intense efforts at the experimental and theoretical levels have been put forth in order to understand the physical origin of charge order and of its interplay with superconductivity \cite{Comin16}. On the one hand, the possible special role played by the tips of the Fermi arcs has been discussed phenomenologically within the context of a magnetically-driven charge-order instability \cite{Davis13,Efetov13,Sachdev13,Meier13,Harrison14,Atkinson15}, where the charge-order wave vectors spanning the hot spots are a manifestation of the pseudogap formation due to charge order, rather than being suggestive of pre-existing Fermi arcs that are unstable to charge order. However, a different proposal attributes the pesudogap phase to the pair-density-wave state \cite{Lee14,Fradkin15}, while charge order only appears in the pesudogap phase as a subsidiary order parameter, and the tips of the Fermi arcs themselves result from an EFS instability around the antinodal region that is distinct from charge order. In particular, the physical origin of charge order has been studied based on the $t$-$J$ model by taking into account the pseudogap effect \cite{Feng16}, where the charge-order state is driven by the pseudogap-induced EFS instability, with the charge-order wave vector corresponding to the straight hot spots on EFS. This study \cite{Feng16} also indicates that charge order is intimately related to pseudogap, and they are a natural consequence of the strong electron correlation in cuprate superconductors. On the other hand, it has been argued that the emergence of charge order in the SC-state is consistent with the picture of the anticorrelation between charge order and superconductivity \cite{Comin16,Hinton16,Tabis14}, i.e., these two order parameters are related, as opposed to simply coexisting and competing. Moreover, a possible common origin of the main instabilities in cuprate superconductors has been suggested, namely, the possibility that the sequence of ordering tendencies (${\bf Q}=0$ order precedes charge order, which in turn precedes the SC order) and the phase diagram as a whole are driven by the strong electron correlation \cite{Comin16,Tabis14}. However, up to now, the finial consensus on the physical origin of charge order and of its interplay with superconductivity has not reached. In this paper, we study the physical origin of charge order and of its interplay with superconductivity in cuprate superconductors within the framework of the kinetic-energy-driven SC mechanism, where the SC-state quasiparticle excitation spectrum is obtained explicitly by taking into account the intertwining between the SC gap and pseudogap. Based on this SC-state quasiparticle excitation spectrum, the main features of charge order in the SC-state of cuprate superconductors are qualitatively reproduced  \cite{Comin16,Comin14,Wu11,Chang12,Ghiringhelli12,Neto14,Fujita14,Croft14,Hucker14,Campi15,Comin15a,Hashimoto15,Peng16,Hinton16}, including the doping dependence of the charge-order wave vector. In particular, we show that as charge order in the normal-state \cite{Feng16}, charge order in the SC-state is also driven by the pseudogap-induced EFS instability, with the charge-order wave vector that is well consistent with the wave vector connecting the straight hot spots on EFS. Although there is a coexistence of charge order and superconductivity below $T_{\rm c}$, this charge order antagonizes superconductivity.

This paper is organized as follows. In Sec. \ref{framework}, we briefly introduces the general formalism of the SC-state quasiparticle spectral function of the $t$-$J$ model in the charge-spin separation fermion-spin representation obtained in terms of the full charge-spin recombination scheme. The quantitative characteristics of the interplay of charge order with superconductivity are discussed in Sec. \ref{Interplay-CO-SC}, where we show that the physical origin of charge order can be interpreted in terms of the formation of the pseudogap by which it means a reconstruction of EFS to form the Fermi pockets, while the intimate interplay between charge order and superconductivity is similar to the intrinsical intertwining between the SC gap and pseuogap. In other words, the pseudogap and charge order have a root in common, they and superconductivity are a natural consequence of the strong electron correlation. Finally, we give a summary and discussions in Sec. \ref{conclusions}.

\section{Formalism}\label{framework}

Superconductivity in cuprate superconductors is a phenomenon in which an assembly of electrons goes into the electron pair-condensed phase as a consequence of the dominance of the interaction between electrons by the exchange of a collective-mode \cite{Carbotte11}. This exchanged collective-mode acts like a bosonic glue to hold the electron pairs together, and is closely related to the SC-state quasiparticle excitations determined by the low-energy electronic structure \cite{Damascelli03,Campuzano04,Zhou07}. On the other hand, the charge-order state is defined as a broken-symmetry state occurring when electrons self-organize into the periodic structures \cite{Comin16}. Therefore charge order and its interplay with superconductivity should be reflected in the low-energy electronic structure. The electronic structure of cuprate superconductors in the SC-state is manifested itself by the energy and momentum dependence of the SC-state quasiparticle excitation spectrum $I({\bf k},\omega)$, which is closely related to the SC-state quasiparticle spectral function as \cite{Damascelli03,Campuzano04,Zhou07},
\begin{eqnarray}\label{ARPES}
I({\bf k},\omega)=|M({\bf k},\omega)|^{2}n_{\rm F}(\omega)A({\bf k},\omega),
\end{eqnarray}
where $M({\bf k},\omega)$ is a matrix element between the initial and final electronic states, and therefore depends on the electron momentum, on the energy and polarization of the income photon. However, following the common practice, the magnitude of $M({\bf k},\omega)$ has been rescaled to the unit in this paper. $n_{\rm F}(\omega)$ is the fermion distribution, while $A({\bf k},\omega)$ is the SC-state quasiparticle spectral function, and is related directly with the imaginary part of the single-electron diagonal Green's function $G({\bf k},\omega)$ as $A({\bf k},\omega)= -2{\rm Im}G({\bf k}, \omega)$. This SC-state quasiparticle excitation spectrum in Eq. (\ref{ARPES}) is measurable via the ARPES technique and can provide the crucial information on EFS, the quasiparticle dispersions, and even the momentum-resolved magnitude of the SC gap \cite{Damascelli03,Campuzano04,Zhou07}.

The quasiparticle excitation spectrum $I({\bf k},\omega)$ in Eq. (\ref{ARPES}) also shows that the microscopic understanding of the physical origin of charge order and of its interplay with superconductivity regains a central role in the context of the essential physics of cuprate superconductors, since the calculation of $I({\bf k},\omega)$ must be performed within the microscopic mechanism of superconductivity. After intensive investigations over more than three decades, now it is widely believed that the $t$-$J$ model on a square lattice contains the essential ingredients to describe superconductivity and the related exotic physics in cuprate superconductors \cite{Anderson87}. Its Hamiltonian is given by,
\begin{eqnarray}\label{tjham}
H=-\sum_{\langle l\hat{a}\rangle\sigma}t_{l\hat{a}}C^{\dagger}_{l\sigma}C_{l+\hat{a}\sigma}+\mu\sum_{l\sigma}C^{\dagger}_{l\sigma}C_{l\sigma} +J\sum_{\langle l \hat{\eta}\rangle}{\bf S}_{l}\cdot {\bf S}_{l+\hat{\eta}},~
\end{eqnarray}
supplemented by the local constraint $\sum_{\sigma}C^{\dagger}_{l\sigma}C_{l\sigma}\leq 1$ to exclude double occupancy, where $C^{\dagger}_{l\sigma}$ ($C_{l\sigma}$) is creation (annihilation) operator for electrons with spin orientation $\sigma=\uparrow,\downarrow$ on lattice site $l$, ${\bf S}_{l}=(S^{\rm x}_{l},S^{\rm y}_{l},S^{\rm z}_{l})$ is spin operator, $\mu$ is the chemical potential, and $J$ is the exchange interaction between the nearest-neighbor (NN) sites $\hat{\eta}$. In this paper, we restrict the hopping of electrons $t_{l\hat{a}}$ to the NN sites $\hat{\eta}$ and next NN sites $\hat{\tau}$ with the amplitudes $t_{l\hat{\eta}}=t$ and $t_{l\hat{\tau}}=-t'$, respectively, while $\langle l\hat{a}\rangle$ means that $l$ runs over all sites, and for each $l$, over its NN sites $\hat{a}=\hat{\eta}$ or next NN sites $\hat{a}=\hat{\tau}$. Hereafter, the parameters are chosen as $t/J=2.5$ and $t'/t=0.3$ as in our previous discussions \cite{Feng16}. The magnitude of $J$ and the lattice constant of the square lattice are the energy and length units, respectively. This $t$-$J$ model (\ref{tjham}) therefore is characterised by a competition between the kinetic energy, which makes electrons itinerant, and the magnetic energy, which makes electrons localized. The strong electron correlation manifests itself by the local constraint of no double electron occupancy, and therefore the crucial requirement is to impose this local constraint properly \cite{Yu92,Feng93,Zhang93,Guillou95,Lee99,Anderson00}. In order to satisfy this local constraint, we employ the fermion-spin formalism \cite{Feng9404,Feng15}, in which the electron operators $C_{l\uparrow}$ and $C_{l\downarrow}$ in the $t$-$J$ model (\ref{tjham}) are replaced by $C_{l\uparrow}=h^{\dagger}_{l\uparrow}S^{-}_{l}$ and $C_{l\downarrow}=h^{\dagger}_{l\downarrow} S^{+}_{l}$, where the spinful fermion operator $h_{l\sigma}=e^{-i\Phi_{l\sigma}}h_{l}$ keeps track of the charge degree of freedom of the constrained electron together with some effects of spin configuration rearrangements due to the presence of the doped hole itself (charge carrier), while the spin operator $S_{l}$ represents the spin degree of freedom of the constrained electron, and then the local constraint of no double occupancy is satisfied at each site. In this fermion-spin representation, the original $t$-$J$ model (\ref{tjham}) can be rewritten as,
\begin{eqnarray}\label{cssham}
H&=&\sum_{\langle l\hat{a}\rangle}t_{l\hat{a}}(h^{\dagger}_{l+\hat{a}\uparrow}h_{l\uparrow}S^{+}_{l}S^{-}_{l+\hat{a}} +h^{\dagger}_{l+\hat{a}\downarrow}h_{l\downarrow}S^{-}_{l}S^{+}_{l+\hat{a}}) \nonumber\\
&-&\mu\sum_{l\sigma}h^{\dagger}_{l\sigma}h_{l\sigma}+J_{{\rm eff}}\sum_{\langle l\hat{\eta}\rangle}{\bf S}_{l}\cdot {\bf S}_{l+\hat{\eta}},~~~~
\end{eqnarray}
where $S^{-}_{l}=S^{\rm x}_{l}-iS^{\rm y}_{l}$ and $S^{+}_{l}=S^{\rm x}_{l}+iS^{\rm y}_{l}$ are the spin-lowering and spin-raising operators for the spin $S=1/2$, respectively, $J_{{\rm eff}}=(1-\delta)^{2}J$, and $\delta=\langle h^{\dagger}_{l\sigma}h_{l\sigma}\rangle=\langle h^{\dagger}_{l}h_{l}\rangle$ is the charge-carrier doping concentration. As an important consequence, the kinetic-energy terms in the $t$-$J$ model (\ref{tjham}) have been transferred as the interaction between charge carriers and spins, and therefore dominates the essential physics of cuprate superconductors.

For a microscopic description of the SC-state of cuprate superconductors, the kinetic-energy-driven SC mechanism has been established based on the $t$-$J$ model (\ref{cssham}) in the fermion-spin representation \cite{Feng15,Feng0306,Feng12}, where the interaction between charge carriers and spins directly from the kinetic energy by the exchange of spin excitations generates the SC-state in the particle-particle channel and pseudogap state in the particle-hole channel, and therefore there is a coexistence of the SC gap and pseudogap below $T_{\rm c}$. However, for the discussions of the electronic state properties, the exact knowledge of the single-electron Green's function (then the quasiparticle spectral function) is of crucial importance \cite{Damascelli03,Campuzano04,Zhou07}. Within the framework of the charge-spin separation \cite{Yu92,Feng93,Zhang93,Guillou95,Lee99,Anderson00}, this single-electron Green's function can be evaluated in terms of the charge-spin recombination. In the conventional charge-spin recombination scheme \cite{Yu92,Feng93,Zhang93,Guillou95,Lee99,Anderson00}, the single-electron Green's function in space-time is a product of the charge-carrier and spin Green's functions, and the resulting Fourier transform is a convolution of the charge-carrier and spin Green's functions. However, in the early days of superconductivity, it has been formally demonstrated that a microscopic theory based on the charge-spin separation can not give a consistent description of EFS and the related quasiparticle excitations in terms of the conventional charge-spin recombination \cite{Feng93,Zhang93,Guillou95}. In this case, within the kinetic-energy-driven SC mechanism, we \cite{Feng15a} have developed recently a full charge-spin recombination scheme to fully recombine a charge carrier and a localized spin into an electron, where it has been realized that the coupling form between the electron quasiparticle and spin excitation is the same as that between the charge-carrier quasiparticle and spin excitation. Based on this full charge-spin recombination, the obtained single-electron Green's function in the normal-state can produce a large EFS with the area that fulfills Luttinger's theorem \cite{Feng16}. Following these discussions, the single-electron diagonal and off-diagonal Green's functions $G({\bf k},\omega)$ and $\Im^{\dagger}({\bf k}, \omega)$ of the $t$-$J$ model (\ref{cssham}) in the fermion-spin representation have been obtained as \cite{Feng15a},
\begin{widetext}
\begin{subequations}\label{t-J-model-EGF}
\begin{eqnarray}
G({\bf k},\omega)&=&{1\over \omega-\varepsilon_{\bf k}-\Sigma_{1}({\bf k},\omega)-[\Sigma_{2}({\bf k},\omega)]^{2}/[\omega+\varepsilon_{\bf k}+ \Sigma_{1}({\bf k},-\omega)]}, \label{DEGF}\\
\Im^{\dagger}({\bf k},\omega)&=&-{\Sigma_{2}({\bf k},\omega)\over [\omega-\varepsilon_{\bf k}-\Sigma_{1}({\bf k},\omega)][\omega+\varepsilon_{\bf k}+ \Sigma_{1}({\bf k},-\omega)]-[\Sigma_{2}({\bf k},\omega)]^{2}},\label{ODEGF}
\end{eqnarray}
\end{subequations}
\end{widetext}
where the bare electron excitation spectrum $\varepsilon_{\bf k}=-Zt\gamma_{\bf k}+Zt'\gamma_{\bf k}'+\mu$, with $\gamma_{\bf k}=({\rm cos}k_{x}+{\rm cos}k_{y})/2$, $\gamma_{\bf k}'= {\rm cos} k_{x}{\rm cos}k_{y}$, and $Z$ is the number of the NN or next NN sites on a square lattice, while the electron self-energies $\Sigma_{1}({\bf k},\omega)$ in the particle-hole channel and $\Sigma_{2}({\bf k},\omega)$ in the particle-particle channel have been evaluated in terms of the full charge-spin recombination, and are given explicitly in Ref. \onlinecite{Feng15a}. In particular, both the electron self-energies $\Sigma_{1}({\bf k},\omega)$ in the particle-hole channel and $\Sigma_{2}({\bf k},\omega)$ in the particle-particle channel are generated by the same interaction of electrons with spin excitations.

Since the electron self-energy $\Sigma_{2}({\bf k},\omega)$ in the particle-particle channel is a coupling of the energy and momentum dependence of the electron pair interaction strength and electron pair order parameter, it is defined as the energy and momentum dependence of the SC gap  \cite{Eliashberg60,Mahan81}, $\bar{\Delta}_{\rm s}({\bf k},\omega)=\Sigma_{2}({\bf k},\omega)$, where following the common practice, the imaginary part of $\Sigma_{2}({\bf k},\omega)$ has been ignored. On the other hand, the electron self-energy $\Sigma_{1}({\bf k},\omega)$ in the particle-hole channel can be divided into two parts: $\Sigma_{1}({\bf k},\omega)={\rm Re}\Sigma_{1}({\bf k},\omega)+i{\rm Im}\Sigma_{1}({\bf k},\omega)$, where ${\rm Re}\Sigma_{1}({\bf k},\omega)$ and ${\rm Im}\Sigma_{1}({\bf k},\omega)$ are the real and imaginary parts, respectively. With the above single-electron diagonal Green's function (\ref{DEGF}), the SC-state quasiparticle spectral function $A({\bf k},\omega)$ now can be obtained explicitly as,
\begin{eqnarray}\label{ESF}
A({\bf k},\omega)={2\Gamma({\bf k},\omega)\over [\omega-\varepsilon_{\bf k}-{\rm Re}\bar{\Sigma}({\bf k},\omega)]^{2}+\Gamma^{2}({\bf k},\omega)},
\end{eqnarray}
and then the dispersion of the quasiparticle state as a function of momentum and therefore EFS itself can be probed by ARPES measurements, where the SC-state quasiparticle scattering rate $\Gamma({\bf k},\omega)$ and the real part of the modified electron self-energy ${\rm Re}\bar{\Sigma}({\bf k}, \omega)$ are given by,
\begin{subequations}\label{MESE}
\begin{eqnarray}
\Gamma({\bf k},\omega)&=&\left |{\rm Im}\Sigma_{1}({\bf k},\omega) \right .\nonumber\\
&-&\left . {\bar{\Delta}^{2}_{\rm s}({\bf k},\omega){\rm Im}\Sigma_{1}({\bf k},-\omega)\over [\omega +\varepsilon_{\bf k}+{\rm Re}\Sigma_{1}({\bf k}, -\omega)]^{2}+[{\rm Im}\Sigma_{1}({\bf k},-\omega)]^{2}}\right |, \nonumber\\
~~~~~\label{EQDSR}\\
{\rm Re}\bar{\Sigma}({\bf k},\omega)&=&{\rm Re}\Sigma_{1}({\bf k},\omega) \nonumber\\
&+&{\bar{\Delta}^{2}_{\rm s}({\bf k},\omega)[\omega+\varepsilon_{\bf k}+{\rm Re}\Sigma_{1}({\bf k},-\omega)]\over [\omega+\varepsilon_{\bf k}+{\rm Re}\Sigma_{1}({\bf k},-\omega)]^{2}+[{\rm Im}\Sigma_{1}({\bf k},-\omega)]^{2}}. \nonumber\\
~~~\label{MRESE}
\end{eqnarray}
\end{subequations}
Substituting this SC-state quasiparticle spectral function $A({\bf k},\omega)$ in Eq. (\ref{ESF}) into Eq. (\ref{ARPES}), we therefore can obtained the SC-state quasiparticle excitation spectrum $I({\bf k},\omega)$. In comparison with the normal-state quasiparticle scattering rate \cite{Feng16} $\Gamma_{\rm N}({\bf k},\omega)$, it is thus shown that there is a additional suppression of the spectral weight below $T_{\rm c}$ due to the SC gap opening.

\section{Interplay between charge order and superconductivity}\label{Interplay-CO-SC}

\begin{figure*}[t!]
\centering
\includegraphics[scale=0.4]{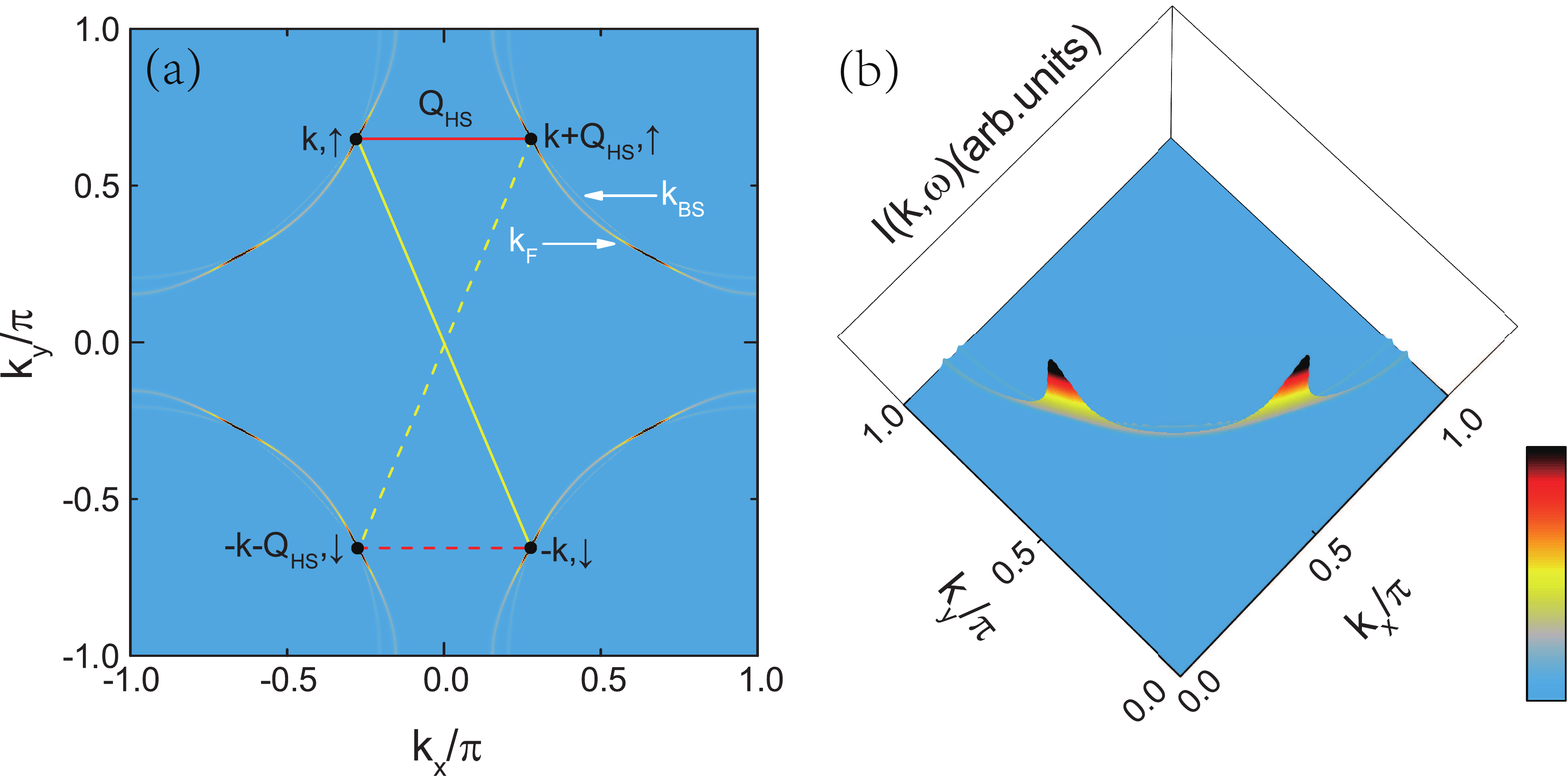}
\caption{(Color online) (a) The map of the quasiparticle excitation spectral intensity $I({\bf k},0)$ and (b) the quasiparticle excitation spectrum $I({\bf k},0)$ in the $[k_{x},k_{y}]$ plane at $\delta=0.12$ with $T=0.002J$ for $t/J=2.5$ and $t'/t=0.3$. The pairing of electrons and holes at ${\bf k}$ and ${\bf k}+{\bf Q}_{\rm HS}$ [red lines in (a)] drives the charge-order formation, whereas the electron pairing at ${\bf k}$ and ${-\bf k}$ states [yellow lines in (a)] is responsible for superconductivity. \label{spectral-maps}}
\end{figure*}

Superconductivity in cuprate superconductors is an instability of the normal-state, and this normal-state from which it emerges over much of the phase diagram is the pseudogap phase \cite{Timusk99,Hufner08}. In this section, we show that the charge-order formation is a natural result of the emergence of the pseudogap, and then in analogy to the interplay between the pseudogap and SC gap, the charge-order correlation is more intimately entangled with superconductivity in cuprate superconductors \cite{Comin16}.

\subsection{Fermi pockets induced by a reconstruction of electron Fermi surface}\label{Fermi-pockets}

EFS of cuprate superconductors can be measured via the ARPES technique \cite{Damascelli03,Campuzano04,Zhou07} and its shape can have deep consequences for the anomalous properties \cite{Gros06}. The nature and topology of EFS in the pseudogap phase of cuprate superconductors has been debated for many years. The early ARPES experimental studies found a large EFS consistent with the band structure calculations \cite{Takahashi89,Campuzano90,Marshall96}. Later, the ARPES measurements with the enhancement of the resolution revealed that the large EFS in the pseudogap phase does not remain intact, but breaks up into the disconnected Fermi arcs \cite{Norman98,Yoshida06,Kanigel07}. Recently, the great improvements in the resolution of the ARPES experimental measurements allowed to resolve additional features in the ARPES spectrum. Among these new achievements is the observation of the Fermi pockets in the pseudogap phase of cuprate superconductors \cite{Yang08,Meng09,Yang11}, with the area of the Fermi pockets that is strongly dependent on the doping concentration \cite{Meng09}. In particular, these Fermi pockets can persist into the SC-state \cite{Meng09}. On the other hand, the charge-order state in cuprate superconductors is closely related to the EFS reconstruction \cite{Comin16,Comin14,Neto14}. This is why the determination of the shape of EFS and the related distribution of the SC-state quasiparticle excitations in the pseudogap phase of cuprate superconductors is believed to be key issue for the understanding of the physical origin of charge order and of its intimate interplay with superconductivity.

The intensity of the SC-state quasiparticle excitation spectrum $I({\bf k},\omega)$ in Eq. (\ref{ARPES}) at zero energy is used to map out the underlying EFS, i.e., the locations of EFS in momentum space is determined directly by \cite{Damascelli03,Campuzano04,Zhou07},
\begin{eqnarray}\label{SC-Fermi-energy}
\varepsilon_{\bf k}+{\rm Re}\bar{\Sigma}_{1}({\bf k},0)=0,
\end{eqnarray}
and then the lifetime of the SC-state quasiparticles at EFS is dominated by the inverse of the SC-state quasiparticle scattering rate $\Gamma({\bf k},0)$ in Eq. (\ref{EQDSR}). For a superconductor, EFS is defined just above $T_{\rm c}$. However, a straightforward calculation shows that $\varepsilon_{\bf k}+{\rm Re}\bar{\Sigma}_{1}({\bf k},0)=0$ in Eq. (\ref{SC-Fermi-energy}) is also equivalent to,
\begin{eqnarray}\label{NS-Fermi-energy}
\varepsilon_{\bf k}+{\rm Re}\Sigma_{1}({\bf k},0)=0,
\end{eqnarray}
which shows that the locations of EFS in momentum space in the SC-state is almost the same as that in the normal-state. This is why we can define operationally {\it EFS} of cuprate superconductors in the SC-state as the contours in momentum space determined from the low-energy spectral weight  \cite{Damascelli03,Campuzano04,Zhou07}. However, the SC quasiparticle scattering rate (then the lifetime of the SC quasiparticle) at EFS has been modified by the SC gap as,
\begin{eqnarray}
\Gamma({\bf k},0)=\left | {\rm Im}\Sigma_{1}({\bf k},0)-{\bar{\Delta}^{2}_{\rm s}({\bf k})\over {\rm Im}\Sigma_{1}({\bf k}, 0)}\right |. \label{MEQDSR}
\end{eqnarray}

In Fig. \ref{spectral-maps}, we plot (a) the map of the SC-state quasiparticle excitation spectral intensity $I({\bf k},0)$ in Eq. (\ref{ARPES}) and (b) the SC-state quasiparticle excitation spectrum $I({\bf k}, 0)$ in the $[k_{x},k_{y}]$ plane at doping $\delta=0.12$ with temperature $T=0.002J$. In the d-wave type SC-state, if the single-particle coherence from the electron self-energy $\Sigma_{1}({\bf k},0)$ in the particle-hole channel is neglected, EFS as the single contour in momentum space is gapped, leading to the four isolated gapless points at the nodes in the momentum space \cite{Gros06}. However, when the single-particle coherence from $\Sigma_{1}({\bf k},0)$ is included as shown in Fig. \ref{spectral-maps}, some unconventional features emerge: (i) the original single-contour EFS in momentum space is split by the electron self-energy $\Sigma_{1}({\bf k}, 0)$ into two contours ${\bf k}_{\rm F}$ and ${\bf k}_{\rm BS}$, respectively, where the redistribution of the low-energy spectral weight of the SC-state quasiparticle excitation spectrum leads to a reconstruction of EFS; (ii) the low-energy spectral weight at the contours ${\bf k}_{\rm F}$ and ${\bf k}_{\rm BS}$ are suppressed by the SC-state quasiparticle scattering rate $\Gamma({\bf k},0)$. However, this suppression is extremely anisotropic. In particular, the low-energy spectral weight around the antinodal region is suppressed heavily, while the low-energy spectral weight around the nodal region is reduced moderately, which leads to that the contours ${\bf k}_{\rm F}$ and ${\bf k}_{\rm BS}$ break up into the disconnected segments around the nodal region; (iii) the contour ${\bf k}_{\rm F}$ intersects the contour ${\bf k}_{\rm BS}$ at the tips of these disconnected segments to form a Fermi pocket, where following the common practice \cite{Yang08,Meng09,Yang11}, the disconnected segment around the nodal region at the contour ${\bf k}_{\rm F}$ is referred to the Fermi arc, and is also defined as the front side of the Fermi pocket, while the other at the contour ${\bf k}_{\rm BS}$ around the nodal region is associated with the back side of the Fermi pocket. Moreover, this Fermi pocket is not symmetrically located in the Brillouin zone (BZ), i.e., it is not centered around $[\pi/2,\pi/2]$. In comparison with the corresponding results of cuprate superconductors in the normal-state \cite{Zhao17}, it is thus shown that the Fermi pockets in cuprate superconductors appeared in the SC-state can persist into the normal-state, and the location, shape and area of the Fermi pockets in the SC-state are almost the same as that in the normal-state. These results are in qualitative agreement with the experimental data obtained by means of the ARPES experimental measurements \cite{Yang08,Meng09,Yang11} and magnetoresistance quantum oscillation \cite{Nicolas07,LeBoeuf07,Sebastian08,Barisic13,Chan16}, where the definitive Fermi pockets in both the SC- and normal-states have been observed.

\subsection{Coexistence of charge order and superconductivity}\label{CO-SC}

In Fig. \ref{spectral-maps}, it is also shown clearly that the part of the spectral weight of the SC-state quasiparticle excitation spectrum at the Fermi arc has been transferred to the back side of the Fermi pocket by the self-energy $\Sigma_{1}({\bf k},0)$, which leads to that although the Fermi arc and back side of the Fermi pocket possess finite spectral weight \cite{Meng09}, the most part of the spectral weight is located around the tips of the Fermi arcs, which in this case coincide with the hot spots on EFS, where the spectral weight has a largest value (see Fig. \ref{spectral-maps}b), reflecting a fact that the most of the SC quasiparticles occupies region around the eight isolated hot spots on EFS, and then these SC quasiparticles around the hot spots contribute effectively to the SC-state quasiparticle scattering process \cite{Comin16}. However, charge order in cuprate superconductors is characterized by the charge-order wave vector, which is just determined by the wave vector connecting the straight hot spots on EFS \cite{Comin16,Comin14,Neto14}. In the present study based on the kinetic-energy-driven SC mechanism, we find that the theoretical result of the SC quasiparticle scattering wave vector between the hot spots on the straight Fermi arcs shown in Fig. \ref{spectral-maps}a at the doping $\delta=0.12$ is $Q_{\rm HS}=0.270$ (hereafter we use the reciprocal units), which is in good agreement with the experimental average value \cite{Comin16,Comin14,Wu11,Chang12,Ghiringhelli12,Neto14,Fujita14,Croft14,Hucker14,Campi15,Comin15a,Hashimoto15,Peng16,Hinton16} of the charge-order wave vector $Q_{\rm CO}\approx 0.265$ oberved in the underdoped cuprate superconductors, indicating that as charge order in the normal-state \cite{Feng16}, charge order in the SC-state is also driven by the EFS instability. Moreover, these results also show that (i) the charge-order-induced reconstruction of EFS into the Fermi pockets is caused by a finite charge-order wave vector $Q_{\rm HS}$ \cite{Nicolas07,LeBoeuf07,Sebastian08,Barisic13,Chan16}; (ii) the charge-order correlation is developed in the normal-state \cite{Feng16}, and can persists into the SC-state, leading to a coexistence of charge order and superconductivity below $T_{\rm c}$. These results are also well consistent with the experimental observations \cite{Comin16,Comin14,Wu11,Chang12,Ghiringhelli12,Neto14,Fujita14,Croft14,Hucker14,Campi15,Comin15a,Hashimoto15,Peng16,Hinton16}. Furthermore, we \cite{Zhao18} have shown that the magnitude of the charge-order wave vector $Q_{\rm HS}$ is also related to the next NN hopping $t'$ of electrons, i.e., it increases with the increase of $t'$, and therefore the experimentally observed differences of the magnitudes of the charge-order wave vector $Q_{\rm CO}$ among the different families of cuprate superconductors at the same doping concentration can be attributed to the different values of $t'$.

In the conventional superconductors \cite{Bardeen57,Schrieffer64}, the SC quasiparticles are the phase-coherent linear superpositions of electrons and holes, while the SC condensate is made up of electron pairs, which are bound-states of two electrons with opposite momenta and spins. However, the charge-order quasiparticles are superpositions of electrons (or holes), and then the charge-order state is likewise a pair condensate, but of electrons and holes, whose net momentum determines the wave-length of charge order \cite{Kohn70,Hinton16}. Within the framework of the kinetic-energy-driven SC mechanism, the structure of the coexistence of superconductivity and charge order in cuprate superconductors is also shown in Fig. \ref{spectral-maps}a, where the pairing of electrons and holes at ${\bf k}$ and ${\bf k}+{\bf Q}_{\rm HS}$ (red lines) separated by the charge-order wave vector $Q_{\rm HS}$ drives the charge-order formation, whereas the electron pairing at ${\bf k}$ and ${-\bf k}$ (yellow lines) states induces superconductivity. In this SC-state with coexisting charge order, the quasiparticles become superpositions of four, rather just two quasiparticle eigenstates \cite{Hinton16}.

\subsection{Doping dependence of charge-order wave vector}

\begin{figure}[h!]
\centering
\includegraphics[scale=0.4]{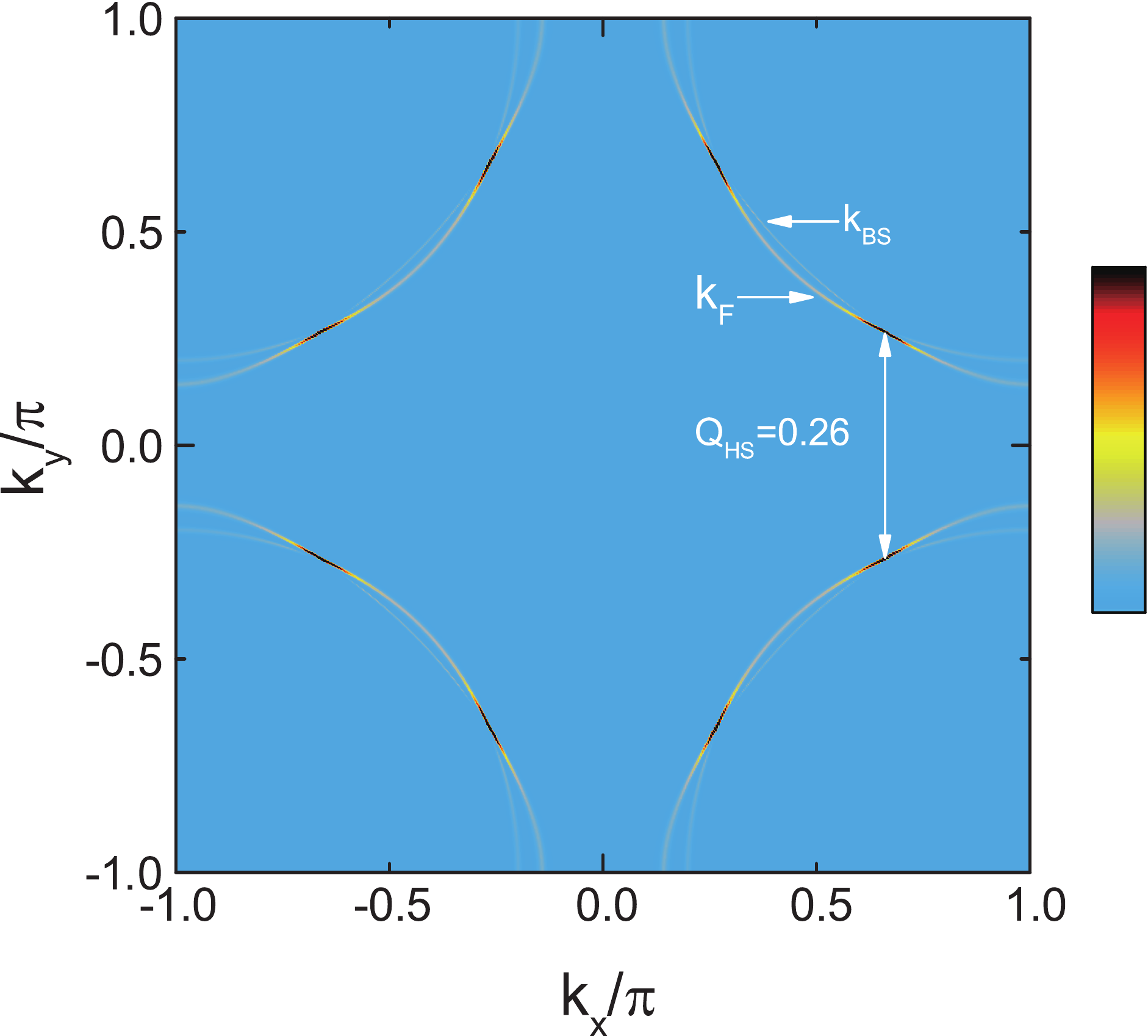}
\caption{(Color online) The map of the quasiparticle excitation spectral intensity $I({\bf k},0)$ at $\delta=0.15$ with $T=0.002J$ for $t/J=2.5$ and $t'/t=0.3$. \label{spectral-maps-doping}}
\end{figure}

\begin{figure}[h!]
\centering
\includegraphics[scale=0.4]{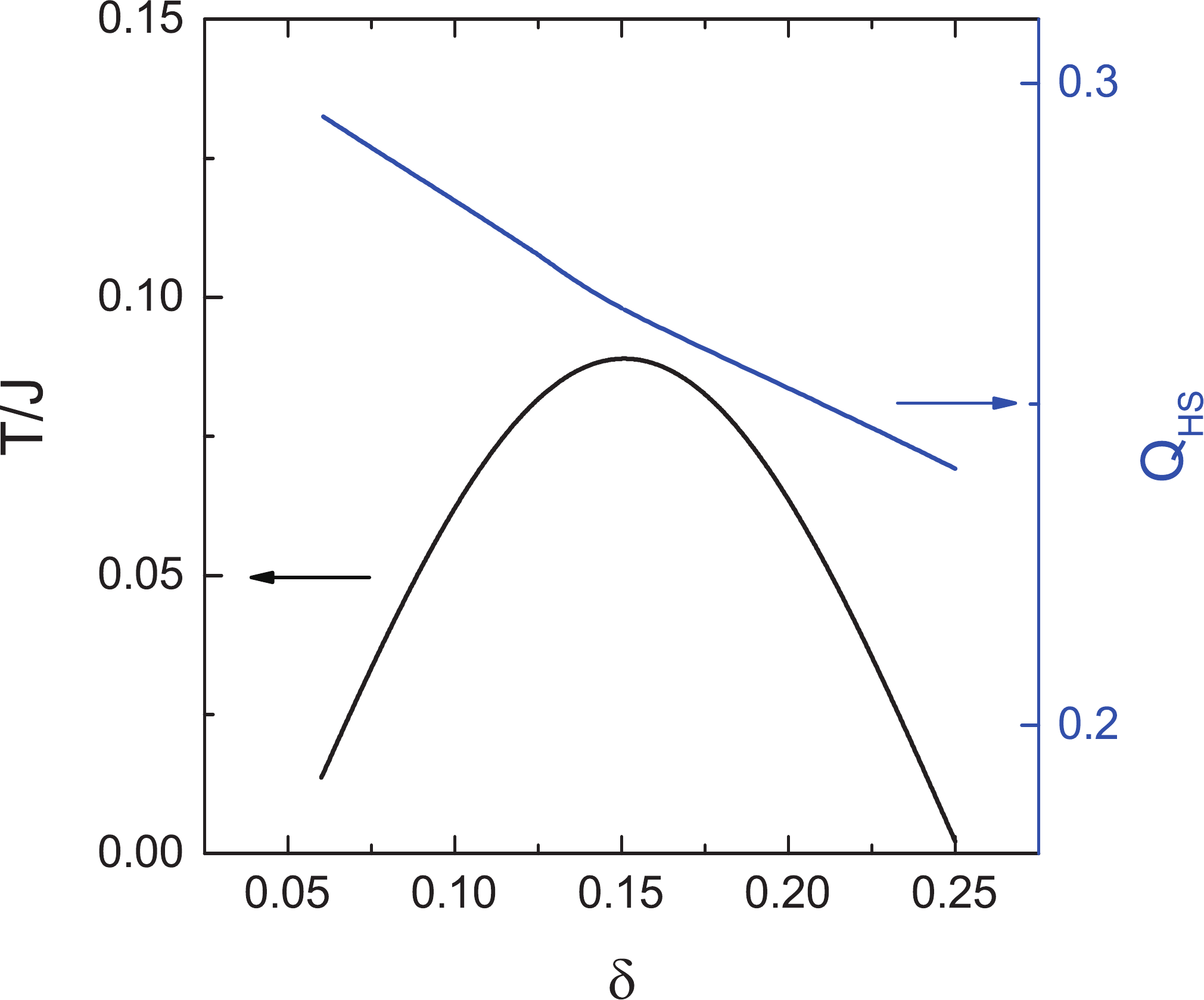}
\caption{(Color online) The charge-order wave vector (blue line) in $T=0.002J$ and superconducting transition temperature $T_{\rm c}$ (black line) for $t/J=2.5$ and $t'/t=0.3$ as a function of doping.  \label{phase-disgram-doping}}
\end{figure}

The above result in Fig. \ref{spectral-maps} indicates that the charge-order wave vector is just corresponding to the straight hot spots on EFS, however, the position of the hot spots is doping dependent. In this case, the charge-order state in cuprate superconductors is characterized not only by the charge-order wave vector, but also by its doping dependence \cite{Comin16}. To address the evolution of the charge-order wave vector with doping, we plot the map of the SC-state quasiparticle excitation spectral intensity $I({\bf k},0)$ at the doping $\delta=0.15$ with $T=0.002J$ in Fig. \ref{spectral-maps-doping}. Comparing it with Fig. \ref{spectral-maps}a for the same set of parameters except for the doping $\delta=0.15$, it is thus shown that with the {\it increase} of doping, the position of the hot spots shifts towards to the antinodes, which leads to that the magnitude of the charge-order wave vector decreases with the increase of doping. To see this doping dependence of the charge-order wave vector more clearly, we have performed a series of calculations for the SC-state quasiparticle excitation spectrum $I({\bf k},0)$ at different doping concentrations, and the result for the extracted charge-order wave vector $Q_{\rm HS}$ (blue line) as a function of doping with $T=0.002J$  is plotted in Fig. \ref{phase-disgram-doping}. For comparison, the result \cite{Feng15,Feng12,Feng15a} of $T_{\rm c}$ (black line) as a function of doping is also replotted in Fig. \ref{phase-disgram-doping}. It is shown clearly that $T_{\rm c}$ has a distinct dome-shaped doping dependence, i.e., it increases with the increase of doping in the underdoped regime, and reaches a maximum in the optimal doping, then decreases with the increase of doping in the overdoped regime. However, in contrast to the case of $T_{\rm c}$ in the underdoped regime, the magnitude of $Q_{\rm HS}$ smoothly decreases with the increase of doping in the underdoped regime, in qualitative agreement with the experimental data \cite{Comin16,Comin14,Wu11,Chang12,Ghiringhelli12,Neto14,Fujita14,Croft14,Hucker14,Campi15,Comin15a,Hashimoto15,Peng16,Hinton16}. Furthermore, in comparison with the corresponding results of the doping dependence of the pseudogap \cite{Timusk99,Hufner08,Feng15,Feng12}, it is thus shown that the behavior of the doping dependence of the charge-order wave vector is very similar to that of the doping dependence of the pseudogap, indicating that the appearance of charge order is closely related to the emergence of the pseudogap.

In spite of a coexistence of charge order and superconductivity below $T_{\rm c}$, above obtained results also show a change in the EFS topology from a single contour in momentum space, where the SC-state quasiparticle excitation spectrum is gapless at the nodes and therefore the SC quasiparticle lifetime on the nodes is infinitely long, to the Fermi pockets, where the SC quasiparticle energies have been heavily renormalized by the electron self-energy $\Sigma_{1}({\bf k},\omega)$ in the particle-hole channel and then they acquire a finite lifetime $\tau({\bf k},0)= \Gamma^{-1}({\bf k},0)$ on the Fermi arc and back side of the Fermi pocket, thereby indicating a reconstruction of EFS caused by the onset of charge order. In this case, the essential physics of the intimate interplay of charge order with superconductivity is closely related to the coexistence and competition between the SC gap and pseudogap below $T_{\rm c}$. This follows a fact \cite{Feng15,Feng12,Feng15a} that within the framework of the kinetic-energy-driven SC mechanism, the pseudogap state in the particle-hole channel is generated by the same electron interaction that also generates SC-state in the article-particle channel. In particular, as we have shown in the previous discussions \cite{Feng15,Feng12,Feng15a}, the electron self-energy $\Sigma_{1}({\bf k},\omega)$ in the particle-hole channel in Eq. (\ref{t-J-model-EGF}) can be also rewritten as,
\begin{eqnarray}\label{PG}
\Sigma_{1}({\bf k},\omega)\approx {[\bar{\Delta}_{\rm PG}({\bf k})]^{2}\over\omega+\varepsilon_{0{\bf k}}},
\end{eqnarray}
with the corresponding energy spectrum $\varepsilon_{0{\bf k}}$ and the momentum dependence of the pseudogap $\bar{\Delta}_{\rm PG}({\bf k})$ can be obtained directly from the electron self-energy $\Sigma_{1}({\bf k},\omega)$ and its antisymmetric part $\Sigma_{\rm 1o} ({\bf k},\omega)$ as $\varepsilon_{0{\bf k}}=-\Sigma_{1}({\bf k},0)/\Sigma_{\rm 1o}({\bf k},0)$ and $\bar{\Delta}_{\rm PG}({\bf k})=\Sigma_{1}({\bf k},0)/ \sqrt{-\Sigma_{\rm 1o}({\bf k},0)}$, respectively, and have been given explicitly in Ref. \onlinecite{Feng15a}. In the present case, this pseudogap $\bar{\Delta}_{\rm PG}({\bf k})$ is identified as being a role of the single-particle coherence by which it means a reconstruction of EFS to form the Fermi pockets. In particular, the corresponding imaginary part of $\Sigma_{1}({\bf k},\omega)$ can be also expressed explicitly in terms of the pseudogap $\bar{\Delta}_{\rm PG}({\bf k})$ as,
\begin{eqnarray}\label{IESE}
{\rm Im}\Sigma_{1}({\bf k},\omega)\approx  2\pi[\bar{\Delta}_{\rm PG}({\bf k})]^{2}\delta(\omega+\varepsilon_{0{\bf k}}),
\end{eqnarray}
which therefore reflects an intimate relation between the SC-state quasiparticle scattering rate in Eq. (\ref{EQDSR}) and pseudogap  \cite{Hashimoto15} $\bar{\Delta}_{\rm PG}({\bf k})$.

Substituting this electron self-energy $\Sigma_{1}({\bf k},\omega)$ in Eq. (\ref{PG}) into Eq. (\ref{t-J-model-EGF}), the single-electron diagonal and off-diagonal Green's functions in Eq. (\ref{t-J-model-EGF}) can be rewritten explicitly as,
\begin{subequations}\label{EGF1}
\begin{eqnarray}
G({\bf k},\omega)&=&\left ({U^{2}_{1{\bf k}}\over\omega-E_{1{\bf k}}}+{V^{2}_{1{\bf k}}\over\omega+E_{1{\bf k}}}\right )\nonumber\\
&+&\left ({U^{2}_{2{\bf k}} \over \omega-E_{2{\bf k}}} +{V^{2}_{2{\bf k}}\over\omega+E_{2{\bf k}}}\right ), \label{DEGF1}\\
\Im^{\dagger}({\bf k},\omega)&=&-{a_{1{\bf k}}\bar{\Delta}_{\rm s}({\bf k})\over 2E_{1{\bf k}}}\left ({1\over\omega-E_{1{\bf k}}}-{1\over\omega + E_{1{\bf k}}}\right )\nonumber\\
&+& {a_{2{\bf k}}\bar{\Delta}_{\rm s}({\bf k})\over 2E_{2{\bf k}}}\left ({1\over\omega-E_{2{\bf k}}}-{1\over\omega+E_{2{\bf k}}} \right ), ~~~ \label{ODEGF1}
\end{eqnarray}
\end{subequations}
where $a_{1{\bf k}}=(E^{2}_{1{\bf k}}-\varepsilon^{2}_{0{\bf k}})/(E^{2}_{1{\bf k}}-E^{2}_{2{\bf k}})$, $a_{2{\bf k}}=(E^{2}_{2{\bf k}}- \varepsilon^{2}_{0{\bf k}})/(E^{2}_{1{\bf k}}-E^{2}_{2{\bf k}})$, and as a natural consequence of the coexistence of the pseudogap and SC gap, the quasiparticles become superpositions of four eigenstates with the corresponding energy eigenvalues $E_{1{\bf k}}$, $-E_{1{\bf k}}$, $E_{2{\bf k}}$, and $-E_{2{\bf k}}$, respectively, where $E_{1{\bf k}}=\sqrt{[K_{1{\bf k}}+K_{2{\bf k}}]/2}$, $E_{2{\bf k}}=\sqrt{[K_{1{\bf k}}-K_{2{\bf k}} ]/2}$, and the kernel functions,
\begin{subequations}
\begin{eqnarray}
K_{1{\bf k}}&=&\varepsilon^{2}_{\bf k}+\varepsilon^{2}_{0{\bf k}}+2\bar{\Delta}^{2}_{\rm PG}({\bf k})+\bar{\Delta}^{2}_{\rm s}({\bf k}),\\
K_{2{\bf k}}&=&\sqrt{(\varepsilon^{2}_{\bf k}-\varepsilon^{2}_{0{\bf k}})b_{1{\bf k}}+4\bar{\Delta}^{2}_{\rm PG}({\bf k})b_{2{\bf k}}+
\bar{\Delta}^{4}_{\rm s}({\bf k})},~~~~~~~
\end{eqnarray}
\end{subequations}
with $b_{1{\bf k}}=\varepsilon^{2}_{\bf k}-\varepsilon^{2}_{0{\bf k}}+2\bar{\Delta}^{2}_{\rm s}({\bf k})$, and $b_{2{\bf k}}=(\varepsilon_{\bf k} - \varepsilon_{0{\bf k}})^{2}+\bar{\Delta}^{2}_{\rm s}({\bf k})$, while the coherence factors of the SC-state with the coexisting pseudogap state are given by,
\begin{subequations}\label{coherence-factors}
\begin{eqnarray}
U^{2}_{1{\bf k}}&=&{1\over 2}\left [a_{1{\bf k}}\left (1+{\varepsilon_{\bf k}\over E_{1{\bf k}}}\right )-a_{3{\bf k}}\left (1+{\varepsilon_{0{\bf k}} \over E_{1{\bf k}}}\right )\right ],\\
V^{2}_{1{\bf k}}&=&{1\over 2}\left [a_{1{\bf k}}\left (1-{\varepsilon_{\bf k}\over E_{1{\bf k}}}\right )-a_{3{\bf k}}\left (1-{\varepsilon_{0{\bf k}} \over E_{1{\bf k}}}\right )\right ],\\
U^{2}_{2{\bf k}}&=&-{1\over 2}\left [a_{2{\bf k}}\left (1+{\varepsilon_{\bf k}\over E_{2{\bf k}}}\right )-a_{3{\bf k}}\left (1+{\varepsilon_{0{\bf k} }\over E_{2{\bf k}}}\right )\right ],\\
V^{2}_{2{\bf k}}&=&-{1\over 2}\left [a_{2{\bf k}}\left (1-{\varepsilon_{\bf k}\over E_{2{\bf k}}}\right )-a_{3{\bf k}}\left (1-{\varepsilon_{0{\bf k} }\over E_{2{\bf k}}}\right )\right ], ~~~~~
\end{eqnarray}
\end{subequations}
and satisfy the sum rule for any wave vector ${\bf k}$: $U^{2}_{1{\bf k}}+V^{2}_{1{\bf k}}+U^{2}_{2{\bf k}}+V^{2}_{2{\bf k}}=1$, where $a_{3{\bf k}}= [\bar{\Delta}_{\rm PG}({\bf k})]^{2}/(E^{2}_{1{\bf k}}-E^{2}_{2{\bf k}})$. In analogy to the normal-state case \cite{Zhao17}, this energy band splitting induced by the pseudogap therefore leads to form two contours ${\bf k}_{\rm F}$ and ${\bf k}_{\rm BS}$ in momentum space as shown in Fig. \ref{spectral-maps}.

\begin{figure}[h!]
\centering
\includegraphics[scale=0.4]{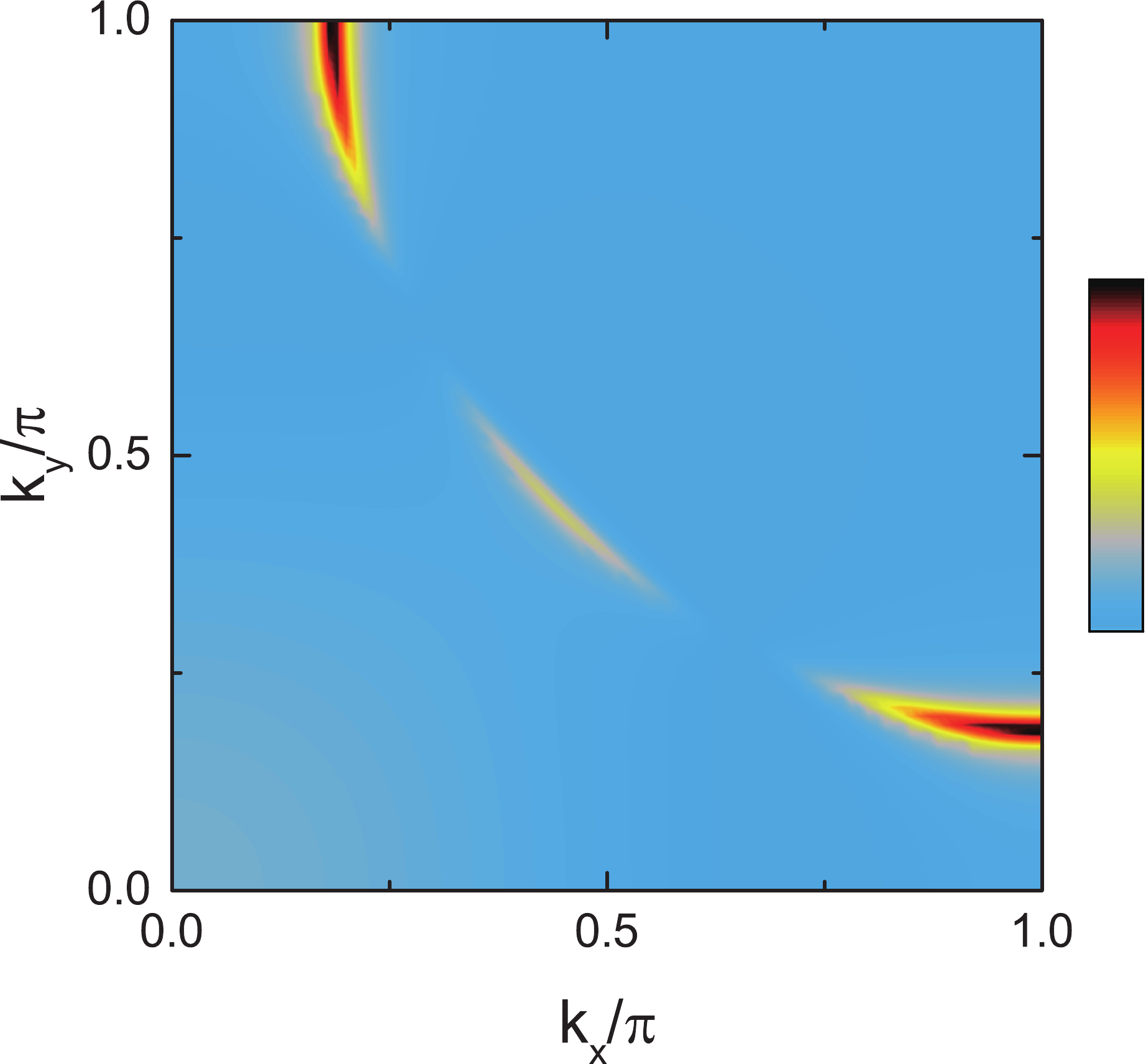}
\caption{(Color online) The map of the quasiparticle scattering rate at $\delta=0.12$ with $T=0.002J$ for $t/J=2.5$ and $t'/t=0.3$. \label{scattering-rate}}
\end{figure}

\begin{figure*}[t!]
\centering
\includegraphics[scale=0.375]{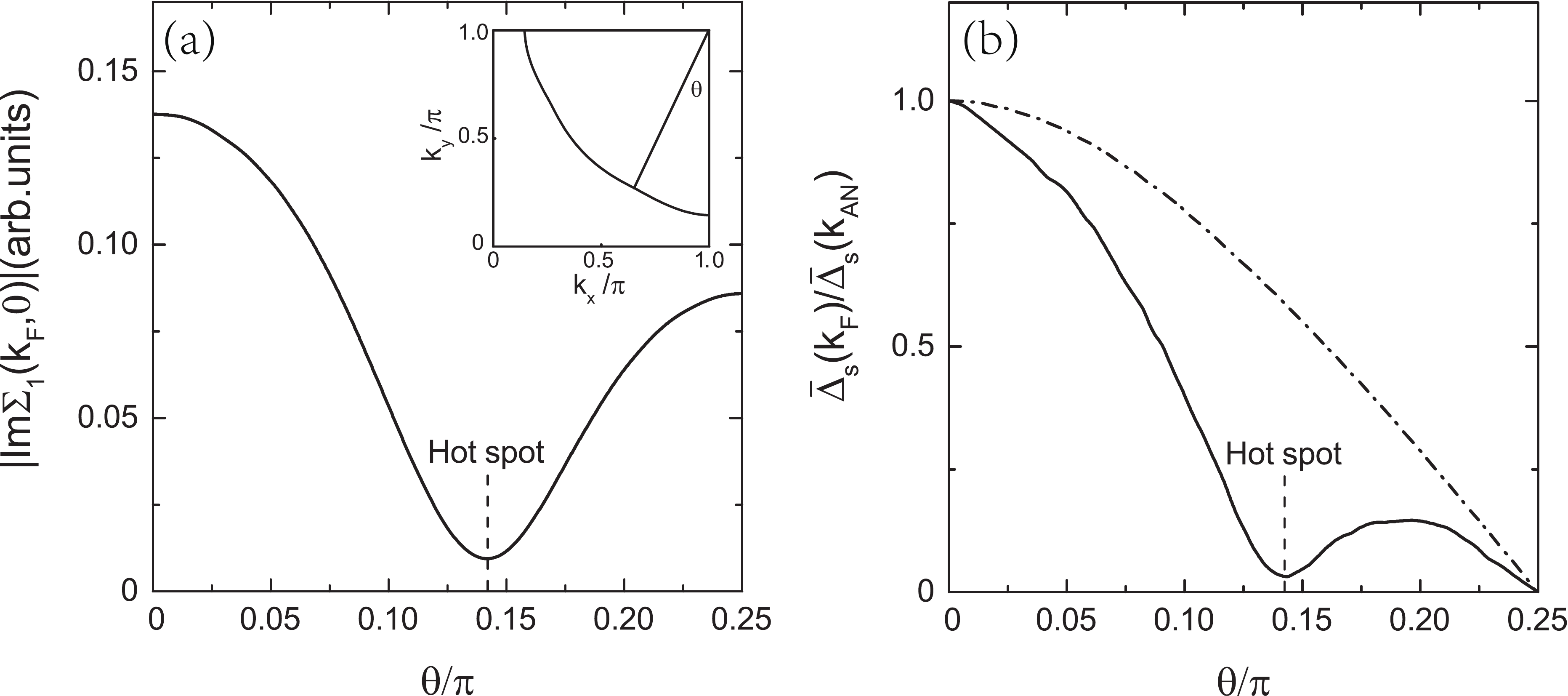}
\caption{The angular dependence of (a) the imaginary part of the electron self-energy in the particle-hole channel and (b) superconducting gap along ${\bf k}_{\rm F}$ from the antinode to the node at $\delta=0.12$ with $T=0.002J$ for $t/J=2.5$ and $t'/t=0.3$. The position of the hot spot is indicated by the dashed vertical line. The dash-dotted line in (b) is obtained from a numerical fit $({\rm cos}k_{x}-{\rm cos}k_{y})/2$. \label{PG-SC-gap}}
\end{figure*}

On the other hand, the contribution to the SC-state quasiparticle excitation spectrum comes from two typical excitations: the electron-hole and the electron pair excitations. In this case, the result in Eq. (\ref{MEQDSR}) also shows that the contribution to the SC-state quasiparticle scattering rate $\Gamma({\bf k},0)$ from the first term ${\rm Im}\Sigma_{1}({\bf k},0)$ of the right-hand side in Eq. (\ref{MEQDSR}) mainly comes from the electron-hole excitations, and is intimately related to the emergence of the pseudogap. This process can therefore persist into the normal-state pseudogap phase, and leads to a number of the anomalous properties \cite{Timusk99,Hufner08}. In particular, the sharp peak structure of the energy and momentum dependence of ${\rm Im}\Sigma_{1}({\bf k},\omega)$ is directly responsible for the remarkable peak-dip-hump structure in the quasiparticle excitation spectrum of cuprate superconductors \cite{DMou17,Gao18}. However, the additional contribution to the SC-state quasiparticle scattering rate $\Gamma({\bf k},0)$ from the second term $\bar{\Delta}^{2}_{\rm s}({\bf k})/{\rm Im}\Sigma_{1}({\bf k},0)$ of the right-hand side in Eq. (\ref{MEQDSR}) originates from the electron pair excitations. This additional process is caused by the SC gap opening, and vanishes in the normal-state. In Fig. \ref{scattering-rate}, we plot the map of the intensity of the SC-state quasiparticle scattering rate $\Gamma({\bf k},0)$ at the doping $\delta=0.12$ with $T=0.002J$. Apparently, $\Gamma({\bf k},0)$ is strong dependence of momentum, reflecting a fact that both the pseudogap $\bar{\Delta}_{\rm PG}({\bf k})$ and SC gap $\bar{\Delta}_{\rm s}({\bf k})$ are extremely anisotropic in momentum space. To see this fact more clearly, we plot the angular dependence of (a) $|{\rm Im}\Sigma_{1}({\bf k}_{\rm F},0)|$ and (b) $\bar{\Delta}_{\rm s}({\bf k}_{\rm F},0)$ along EFS from the antinode to the node at the doping $\delta=0.12$ with $T=0.002J$ in Fig. \ref{PG-SC-gap}. For comparison, the corresponding result (dash-dotted line) of the monotonic d-wave form $({\rm cos}k_{x}-{\rm cos} k_{y})/2$ along EFS is also presented in Fig. \ref{PG-SC-gap}b. These results in Fig. \ref{PG-SC-gap} show that both $|{\rm Im}\Sigma_{1}({\bf k}_{\rm F},0)|$ [then the pseudogap $\bar{\Delta}_{\rm PG} ({\bf k}_{\rm F})$] and $\bar{\Delta}_{\rm s}({\bf k}_{\rm F})$ have a strong angular dependence. On the one hand, $|{\rm Im}\Sigma_{1}({\bf k}_{\rm F},0)|$ exhibits the largest value around the antinode ${\bf k}_{\rm AN}$. However, the actual minimum of $|{\rm Im}\Sigma_{1}({\bf k}_{\rm F},0)|$ does not appear around the node ${\bf k}_{\rm N}$, but locates exactly at the hot spot ${\bf k}_{\rm HS}$, where $|{\rm Im}\Sigma_{1}({\bf k}_{\rm HS},0)|\sim 0$. In particular, the magnitude of $|{\rm Im}\Sigma_{1}({\bf k}_{\rm F},0)|$ around the antinode is larger than that around the node. On the other hand, although the SC gap $\bar{\Delta}_{\rm s}({\bf k}_{\rm F})$ has a dominated d-wave symmetry with the actual maximum at the antinode and the actual minimum [$\bar{\Delta}_{\rm s}({\bf k}_{\rm N})=0$] at the node, it exhibits an unusual momentum dependence around the hot spot region with the anomalously small value, which leads to that the angular variation of $\bar{\Delta}_{\rm s}({\bf k}_{\rm F})$ on EFS can not be described by a monotonic d-wave SC gap form $({\rm cos}k_{x}-{\rm cos}k_{y})/2$, i.e., it cannot be fit by a simple $({\rm cos}k_{x}-{\rm cos}k_{y})/2$ form except for the the nodal and antinodal regions \cite{Mesot99} as shown in Fig. \ref{PG-SC-gap}b.

\begin{figure}[h!]
\centering
\includegraphics[scale=0.4]{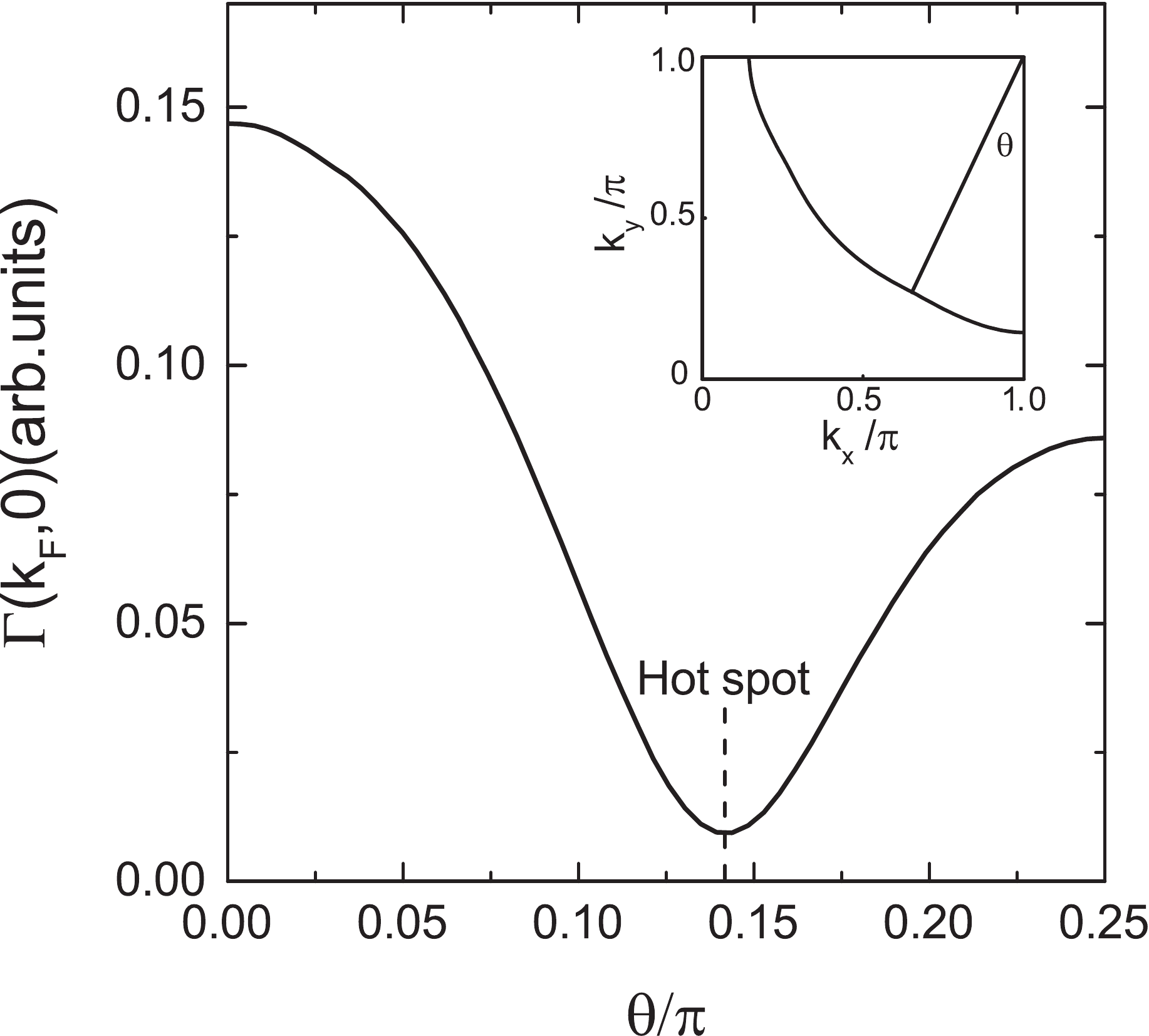}
\caption{The angular dependence of the quasiparticle scattering rate along ${\bf k}_{\rm F}$ from the antinode to the node at $\delta=0.12$ with $T=0.002J$ for $t/J=2.5$ and $t'/t=0.3$.  \label{scattering-rate-1}}
\end{figure}

The combination of both the results of $|{\rm Im}\Sigma_{1}({\bf k}_{\rm F},0)|$ and $\bar{\Delta}_{\rm s}({\bf k}_{\rm F})$ shown in Fig. \ref{PG-SC-gap}, the angular dependence of the SC-state quasiparticle scattering rate $\Gamma({\bf k}_{\rm F},0)$ on EFS can therefore be obtained, and the result is plotted in Fig. \ref{scattering-rate-1}. Moreover, we have also calculated the angular dependence of the SC-state quasiparticle scattering rate $\Gamma({\bf k}_{\rm BS},0)$ along the back side of the Fermi pocket from the antinode to the node, and found that its behavior is very similar to that of $\Gamma({\bf k}_{\rm F},0)$. The result in Fig. \ref{scattering-rate-1} therefore shows clearly that the essential property of the angular dependence of $\Gamma({\bf k}_{\rm F},0)$ is mainly dominated by the pseudogap, i.e., the SC-state quasiparticle scattering rate $\Gamma({\bf k}_{\rm F},0)$ has almost the same momentum dependence as $|{\rm Im}\Sigma_{1}({\bf k}_{\rm F},0)|$. In particular, $\Gamma({\bf k}_{\rm F}, 0)$ exhibits the strongest scattering at the antinodes, which is consistent with the experimental results \cite{Comin14,Neto14,Valla00,Kaminski05,Shi08,Sassa11}, where the strongest quasiparticle scattering appeared at the antinodes has been also widely observed in cuprate superconductors. However, the weakest quasiparticle scattering does not take place at the nodes, but occurs exactly at the hot spots ${\bf k}_{\rm HS}$, where the energy spectra $\varepsilon_{0{\bf k}_{\rm HS}}\approx-\varepsilon_{{\bf k}_{\rm HS}}$, $\bar{\Delta}_{\rm PG} ({\bf k}_{\rm HS})$, and $\bar{\Delta}_{\rm s}({\bf k}_{\rm HS})$ have anomalously small values, and then all the energy bands for the SC-state quasiparticle excitation spectra on the contours ${\bf k}_{\rm F}$ and ${\bf k}_{\rm BS}$ converge on the hot spots of EFS. This extremely anisotropic momentum dependence of the SC-state quasiparticle scattering rates (then the pseudogap) on ${\bf k}_{\rm F}$ and ${\bf k}_{\rm BS}$ therefore suppresses heavily the spectral weight of the SC-state quasiparticle excitation spectrum on the contours ${\bf k}_{\rm F}$ and ${\bf k}_{\rm BS}$ in the antinodal region, and reduces modestly the spectral weight around the nodal region. This is also why the tips of these disconnected segments on ${\bf k}_{\rm F}$ and ${\bf k}_{\rm BS}$ assemble on the hot spots to form a Fermi pocket in the SC-state around the nodal region. At the same time, this EFS instability therefore drives charge order with the charge-order wave vector connecting the parallel hot spots on EFS, as charge order driven by the EFS instability in the normal-state case \cite{Feng16}. In other words, the formation of the Fermi pockets (then the charge-order state) is a natural consequence of the extremely anisotropic momentum dependence of the pseudogap (then the quasiparticle scattering rates) originated from the electron self-energy due to the interaction between electrons by the exchange of spin excitations. Furthermore, the position of the hot spots is doping dependent, and shifts towards to the antinodes when doping is increased, which therefore leads to that the charge-order wave vector decreases with the increase of doping \cite{Comin16,Comin14}. In the normal-state [$\bar{\Delta}_{\rm s}({\bf k},\omega) =0$], the SC-state quasiparticle scattering rate $\Gamma({\bf k},\omega)$ in Eq. (\ref{EQDSR}) is reduced as the normal-state quasiparticle scattering rate $\Gamma_{\rm NS}({\bf k},\omega)=|{\rm Im}\Sigma_{1}({\bf k},\omega)|$. Since the essential properties of the angular dependence of $\Gamma({\bf k}_{\rm F},0)$ is dominated by the pseudogap $\bar{\Delta}_{\rm PG}({\bf k})$, these Fermi pockets persist into the normal-state \cite{Meng09,Zhao17}, and then charge order driven by the EFS instability emerges in the normal-state \cite{Feng16}.

In our previous studies \cite{Feng15a,Feng12,Feng15}, we have shown that (i) there is a coexistence of the SC gap and pseudogap below $T_{\rm c}$; (ii) the pseudogap is directly related to the single-particle coherent weight, and then the SC-state in the kinetic-energy-driven SC mechanism is controlled by both the SC gap and single-particle coherence; (iii) however, this single-particle coherence competes strongly with the electron pairing state, which leads $T_{\rm c}$ in cuprate superconductors to be reduced to lower temperatures, indicating that the pseudogap has a competitive role in engendering superconductivity. With these previous studies of the intertwining between the pseudogap and SC gap, our present results of the connection of charge order and pseudogap therefore show that (i) the reconstruction of EFS and the related charge-order state in the pseudogap phase can be attributed to the emergence of the pseudogap; (ii) as the role played by the pseudogap, although charge order coexists with superconductivity below $T_{\rm c}$, this charge order displays the analogous competition with superconductivity; (iii) the Fermi pocket, charge order, and pseudogap are intimately related, i.e., there is a common origin for the Fermi pocket, charge order, and pseudogap, they and superconductivity are a natural consequence of the strong electron correlation in cuprate superconductors.

\subsection{Quantitative connection between electronic structure and collective response of electron density}\label{charge-order}

Now we turn our attention to the quantitative connection between the collective response of the electron density and low-energy electronic structure. The collective response of the electron density, as manifested by the SC-state dynamical charge structure factor $C({\bf k}, \omega)$, is closely related to the imaginary part of the SC-state quasiparticle density-density correlation function,
\begin{eqnarray}\label{DCSF}
C({\bf k},\omega)=-{1\over\omega}{\rm Im}\tilde{\Pi}_{\bf c}({\bf k},\omega),
\end{eqnarray}
with the SC-state quasiparticle density-density correlation function $\tilde{\Pi}_{\bf c}({\bf k},\omega)$ that is defined as,
\begin{eqnarray}\label{EDDCF}
\tilde{\Pi}_{\bf c}({\bf R}_{l}-{\bf R}_{l'},t-t')=\langle\langle T\rho({\bf R}_{l},t)\rho({\bf R}_{l'},t')\rangle\rangle,
\end{eqnarray}
where the $\rho({\bf R}_{l},t)$ is the density operator, and can be expressed explicitly as,
\begin{eqnarray}\label{EDO}
\rho({\bf R}_{l})=e\sum_{\sigma}C^{\dagger}_{l\sigma}C_{l\sigma}.
\end{eqnarray}
Substituting this density operator (\ref{EDO}) into Eqs. (\ref{EDDCF}) and (\ref{DCSF}), the SC-state dynamical charge structure factor $C({\bf k}, \omega)$ thus can be obtained explicitly in terms of the SC-state quasiparticle spectral functions as,
\begin{widetext}
\begin{eqnarray}\label{DCSF1}
C({\bf k},\omega)&=&2e^{2}{1\over N}\sum_{\bf q}\int^{\infty}_{-\infty}{d\omega'\over 2\pi}[A({\bf q}+{\bf k},\omega'+\omega)A({\bf q},\omega')- A_{\Im}({\bf q}+{\bf k},\omega'+\omega)A_{\Im}({\bf q},\omega')]{n_{\rm F}(\omega'+\omega)-n_{\rm F}(\omega')\over\omega}, ~~~~~~~
\end{eqnarray}
\end{widetext}
where $A_{\Im}({\bf k},\omega)=-2{\rm Im}\Im^{\dagger}({\bf k},\omega)$. It should be noted that in the calculation of the above SC-state dynamical charge structure factor, the vertex correction has been dropped, since it has been shown that the vertex correction is negligibly small in the calculation of the density-density correlation of cuprate superconductors \cite{Lin12,Jing17}.

\begin{figure}[h!]
\centering
\includegraphics[scale=0.4]{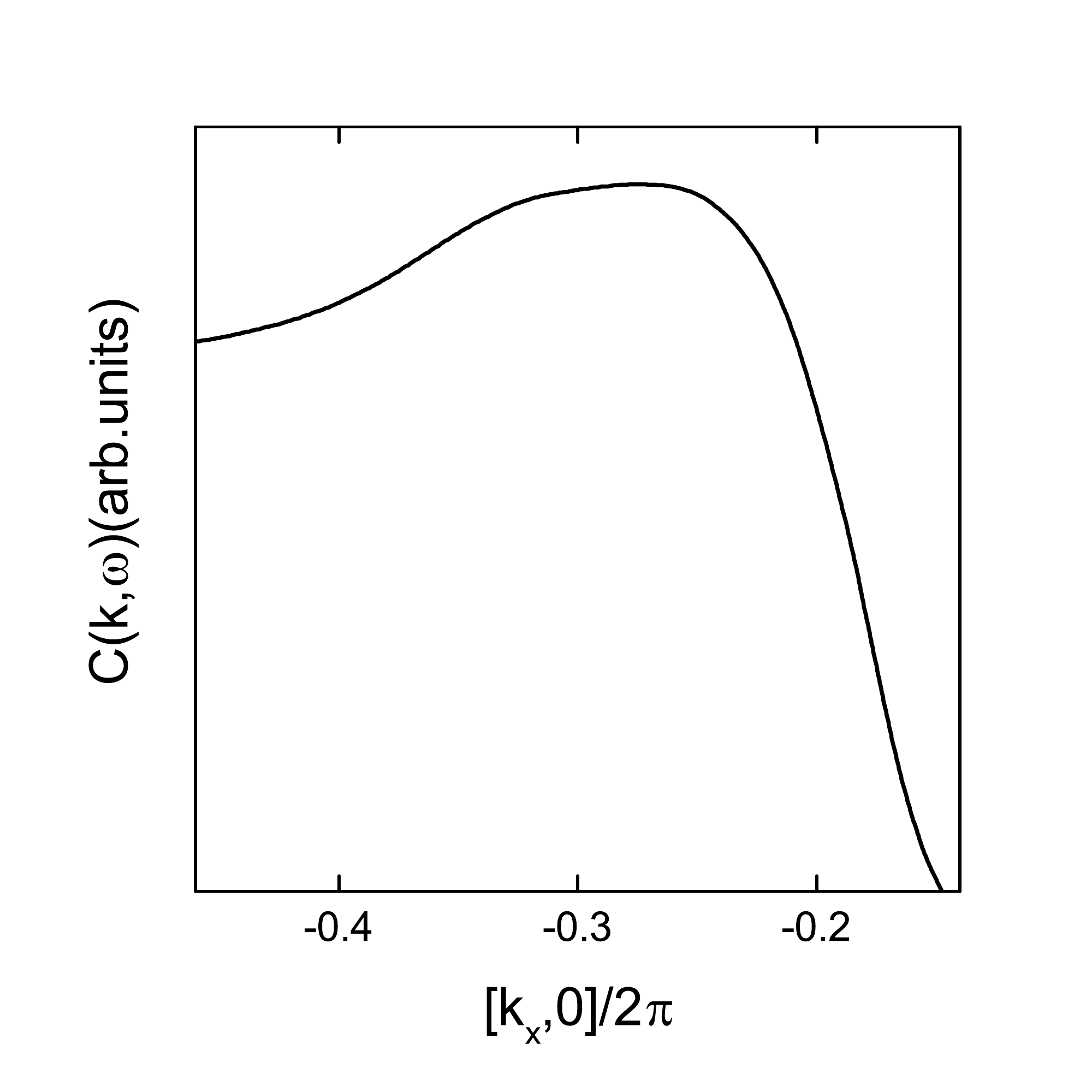}
\caption{Dynamical charge structure factor along the ${\bf k}= [0,0]$ to ${\bf k}=[\pi,0]$ direction of the Brillouin zone at $\omega=6.8J$ and $\delta=0.12$ with $T=0.002J$ for $t/J=2.5$ and $t'/t=0.3$. \label{DCFS}}
\end{figure}

\begin{figure}[h!]
\centering
\includegraphics[scale=0.4]{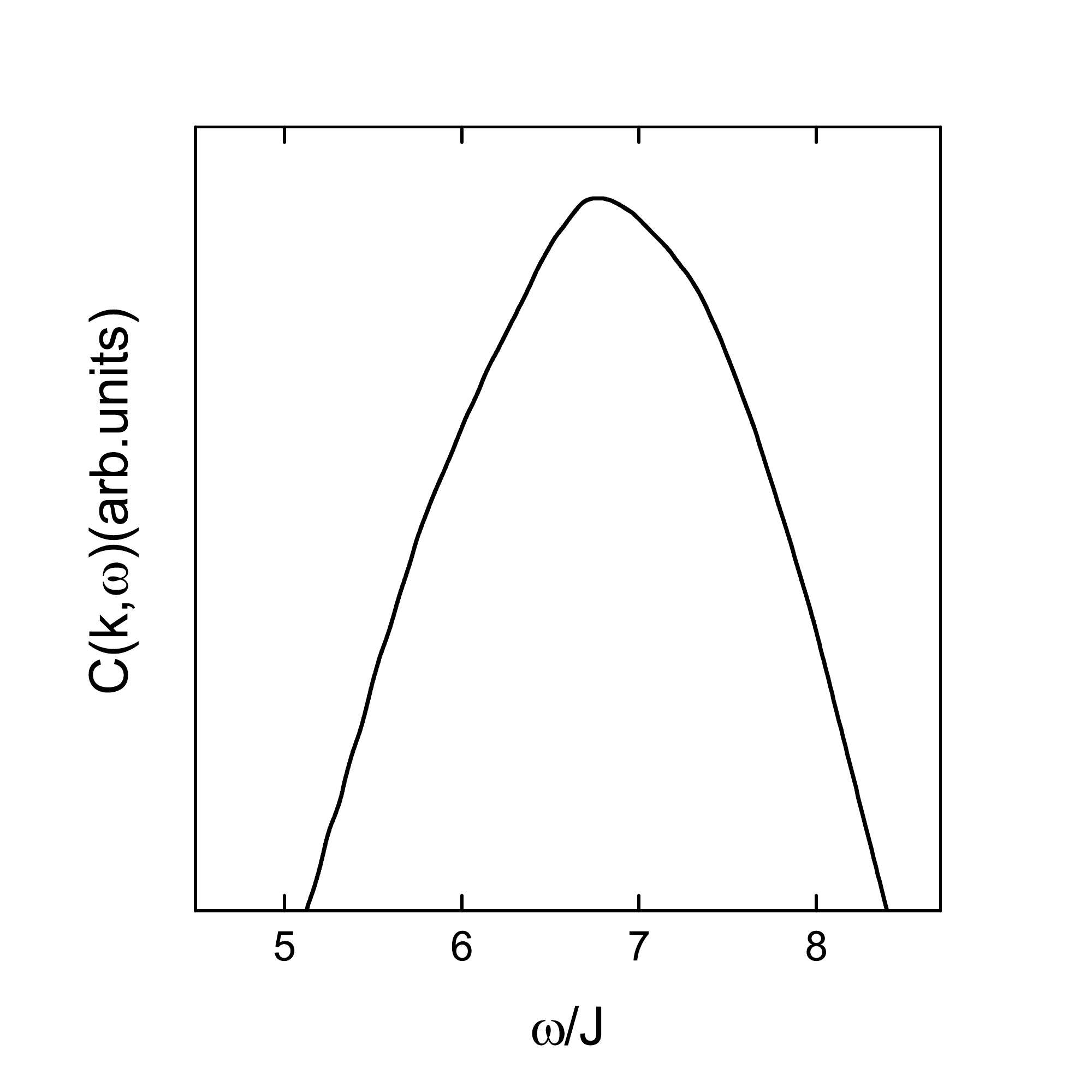}
\caption{Dynamical charge structure factor as a function of energy in the charge-order wave vector $Q_{\rm CD}=0.27$ at $\delta=0.12$ with $T=0.002J$ for $t/J=2.5$ and $t'/t=0.3$. \label{DCFS-energy}}
\end{figure}

To explore the global feature of charge order in the SC-state, we have mapped the SC-state dynamical charge structure factor (\ref{DCSF1}) in the $[k_{x},k_{y}]$ plane, and find that there are four resonance peaks, where the charge-order propagates only along the parallel directions $[\pm Q_{\rm HS},0]$ and $[0,\pm Q_{\rm HS}]$ of BZ, in good agreement with the experimental observations \cite{Comin16,Comin14,Wu11,Chang12,Ghiringhelli12,Neto14,Fujita14,Croft14,Hucker14,Campi15,Comin15a,Hashimoto15,Peng16,Hinton16}. To see the parallel resonance peaks around the wave vector $Q_{\rm HS}$ clearly, we plot the result of $C({\bf k},\omega)$ along the ${\bf k}=[0,0]$ to ${\bf k}= [\pi,0]$ direction of BZ at the energy $\omega=6.8J$ and the doping $\delta=0.12$ with $T=0.002J$ in Fig. \ref{DCFS}, where a resonance peak emerges in the wave vector $Q_{\rm CO}\approx 0.27$. In particular, this characteristic wave vector $Q_{\rm CO}\approx 0.27$ is the exactly same with the charge-order wave $Q_{\rm HS}=0.270$ determined from the corresponding to the straight hot spots on EFS in Sec. \ref{CO-SC}, indicating a quantitative connection between the electronic structure and collective response of electron density. Since the intensity of the resonance peak is determined by damping, it is thus fully understandable that the intensity of the resonance peak is suppressed as the temperature is increased \cite{Comin16,Comin14,Wu11,Chang12,Ghiringhelli12,Neto14,Fujita14,Croft14,Hucker14,Campi15,Comin15a,Hashimoto15,Peng16,Hinton16}. To further verify this resonance peak that can be identified as the presence of charge ordering, we plot $C(Q_{\rm CO},\omega)$ as a function of energy in the wave vector $Q_{\rm CO}=0.27$ at the doping $\delta=0.12$ with $T=0.002J$ in Fig. \ref{DCFS-energy}, where the energy is tuned away from the resonance energy, the sharp peak in the wave vector $Q_{\rm CO}=0.27$ is suppressed heavily, and then eventually disappears, also in qualitative agreement with the experimental observation from the RXS measurements \cite{Comin16,Comin14,Wu11,Chang12,Ghiringhelli12,Neto14,Fujita14,Croft14,Hucker14,Campi15,Comin15a,Hashimoto15,Peng16,Hinton16}. This suppression of the resonance peak by tuning energy away from the resonance energy therefore verifies the presence of charge order as the collective response of the electron density in the SC-state of cuprate superconductors.

The above obtained results therefore confirm a quantitative connection between the low-energy electronic structure and the collective response of electron density in cuprate superconductors \cite{Comin16,Comin14,Neto14}. The SC-state dynamical charge structure factor $C({\bf k},\omega)$ in Eq. (\ref{DCSF1}) is obtained in terms of the SC-state quasiparticle spectral functions, in other words, the essential behavior of the collective response of the electron density in the SC-state of cuprate superconductors is mainly determined by the SC-state quasiparticle spectral functions [then the single-electron diagonal and off-diagonal Green's functions in Eq. (\ref{EGF1}) and the related pesudogap in Eq. (\ref{PG})]. However, the single-electron diagonal and off-diagonal Green's functions (\ref{EGF1}) and the related dispersions of the SC-state quasiparticle excitation energies $E_{1{\bf k}}$ and $E_{2{\bf k}}$ can be also reproduced qualitatively by a phenomenological Hamiltonian,
\begin{eqnarray}\label{CO-model}
H_{\rm CO}&=&\sum_{{\bf k}\sigma}\varepsilon_{\bf k}C^{\dagger}_{{\bf k}\sigma}C_{{\bf k}\sigma}-\sum_{\bf k}\bar{\Delta}_{\rm s}({\bf k}) (C^{\dagger}_{{\bf k}\uparrow}C^{\dagger}_{-{\bf k}\downarrow}+C_{-{\bf k}\downarrow}C_{{\bf k} \uparrow})\nonumber\\
&-&\sum_{{\bf k}\sigma}\varepsilon_{0{\bf k}} C^{\dagger}_{{\bf k}+{\bf Q}_{\rm HS}\sigma}C_{{\bf k}+{\bf Q}_{\rm HS}\sigma}\nonumber\\
&+&\sum_{{\bf k}\sigma} \bar{\Delta}_{\rm PG}({\bf k})(C^{\dagger}_{{\bf k}+{\bf Q}_{\rm HS}\sigma}C_{{\bf k}\sigma}+C^{\dagger}_{{\bf k} \sigma}C_{{\bf k}+{\bf Q}_{\rm HS} \sigma}), ~~~~~~~~~
\end{eqnarray}
while such type Hamiltonian has been usually employed to phenomenologically discuss the physical origin of charge order and of its interplay with superconductivity in cuprate superconductors \cite{Davis13,Efetov13,Sachdev13,Meier13,Harrison14}, where charge order induces a reconstruction of EFS to form the Fermi pockets, in qualitative agreement with the experimental data. This is why our present study based on the kinetic-energy-driven SC mechanism can give a consistent description of the interplay between charge order and superconductivity in cuprate superconductors.

\section{Conclusions}\label{conclusions}

In conclusion, within the framework of the kinetic-energy-driven SC mechanism, we have discussed the physical origin of charge order and of its interplay with superconductivity in cuprate superconductors by taking into account the intertwining between the pseudogap and SC gap. In particular, we show that the pseudogap-induced EFS reconstruction generates the formation of the Fermi pockets around the nodal region. However, the distribution of the spectral weight of the SC-state quasiparticle excitation spectrum on the Fermi arc and back side of the Fermi pocket is extremely anisotropic, where the most of the SC quasiparticles occupies region around the hot spots on EFS. As charge order in the normal-state, this EFS instability drives the charge-order correlation in the SC-state, with the charge-order wave vector connecting the straight hot spots on EFS. As a natural consequence of a doped Mott insulator, this charge-order state is doping dependent, with the magnitude of the charge-order wave vector that decreases with the increase of doping. Although charge order appears to be a phenomenon that coexists with superconductivity below $T_{\rm c}$, this charge order strongly competes with superconductivity. The results from the SC-state dynamical charge structure factor indicate the existence of a quantitative connection between the low-energy electronic structure and collective response of the electron density. Combined with the previous results in the normal-state \cite{Feng16,Zhao17}, the theory also shows that the pseudogap and charge order are both consequences of the same physics, they and superconductivity are a natural result of the strong electron correlation in cuprate superconductors.

\section*{Acknowledgements}

The authors would like to thank Professor Yongjun Wang for helpful discussions. DG, YL, YM, and SF are supported by the National Key Research and Development Program of China under Grant No. 2016YFA0300304, and the National Natural Science Foundation of China (NSFC) under Grant Nos. 11574032 and 11734002. HZ is supported by NSFC under Grant No.11547034.

\end{document}